\documentstyle[12pt,epsf]{article}

\newcommand{\bea}{\begin{eqnarray}}
\newcommand{\eea}{\end{eqnarray}}
\newcommand{\be}{\begin{equation}}
\newcommand{\ee}{\begin{equation}}

\begin{document}

\parindent 0pt
\renewcommand{\thefootnote}{\fnsymbol{footnote}}

\begin{flushright}
\texttt{hep-ph/0205268}\\
IC/2002/34\\
SINP/TNP/02-18\\
\end{flushright}

\vskip 30pt

\begin{center}
{\large \bf Gauge unification in 5-D $SU(5)$ model with orbifold breaking
of GUT symmetry}
\vspace{.5in}

{\bf Biswajoy Brahmachari$^{a,c}$
\footnote{e-mail: biswajoy@theory.saha.ernet.in} 
and 
Amitava Raychaudhuri$^{b,c}$
\footnote{e-mail: amitava@cubmb.ernet.in} 
}

\vskip 1cm

(a) Theoretical Physics Group, Saha Institute of Nuclear Physics \\
AF/1 Bidhannagar, Kolkata 700064, India \\
\vskip .5cm

(b) Department of Physics, University of Calcutta \\
92, Acharya Prafulla Chandra Road, Kolkata 700009, India
\vskip .5 cm

(c) Abdus Salam International Centre for
Theoretical Physics \\
Strada Costiera 11, 34014 Trieste, Italy\\
\vskip 1cm
\underbar{ABSTRACT}
\end{center}

We consider a 5-dimensional $SU(5)$ model wherein the symmetry is broken
to the 4-dimensional Standard Model by compactification of the 5th
dimension on an $S^1/(Z_2 \times Z^\prime_2)$ orbifold. We identify the
members of all $SU(5)$ representations upto {\bf 75} which have zero modes. We
examine how these light scalars affect gauge coupling unification assuming
a single intermediate scale and present several acceptable solutions. 
The 5-D compactification scale coincides with the unification scale of 
gauge couplings and is determined via this renormalization group analysis.
When $SO(10)$ is considered as the GUT group there are only two solutions, so 
long as a few low dimensional scalar multiplets upto {\bf 126} are included.

\newpage
\setcounter{footnote}{0}
\renewcommand{\thefootnote}{\arabic{footnote}}

\section{Introduction}

The $SU(5)$ model\cite{su5} unifies the strong, weak, and
electromagnetic interactions in the smallest simple group. It has many
other attractive features which are well recognized. But it suffers
from the following two major difficulties which are actually generic
to the idea of grand unification\cite{guts} itself.  (i) Because
quarks and leptons reside in unified multiplets and there are B- and
L-violating interactions, gauge boson exchanges can result in proton
decay\cite{pd}.  If these gauge bosons are appropriately heavy, the decay rate
will be very small. Their masses, in the usual formulation, are,
however, not arbitrary but rather determined by the scale where the
different gauge couplings unify.  The proton decay lifetime is
therefore a robust prediction of the model. No experimental signature
of proton decay\cite{pdexp} has been found yet and the model is
disfavoured. More complicated unification models involving several
intermediate mass-scales can evade this problem\cite{gaugeb}.  (ii) 
The low energy Higgs doublet, responsible for electroweak breaking, is 
embedded in a {\bf 5} representation of $SU(5)$.  The other members of this 
multiplet are color triplet scalars which must have a mass near the 
unification scale -- since no such scalars have been observed at the 
electroweak scale. This leads to an unnatural mass splitting among the 
members of the same $SU(5)$ multiplet.  This is termed the double-triplet
splitting problem\cite{dt}.

These two unwelcome features of the $SU(5)$ model can be tackled in an
elegant way if unified $SU(5)$ symmetry exists in a 5-D world. Low
energy 4-D $SU(3)_c \times SU(2)_L \times U(1)_Y$ symmetry is
recovered when the extra dimension is compactified on a $S^1 / (Z_2
\times Z^\prime_2)$ orbifold\cite{szz}. This situation is realized
when space-time is considered to be factorized into a product of 4D
Minkowski space-time $M^4$ and the orbifold $S^1 / (Z_2 \times
Z^\prime_2)$. The coordinate system consists of
$x^\mu=(x^0,x^1,x^2,x^3)$ and $y=x^5$. There are two distinct 4-D
branes; one at $y = 0$ and another at $y =\pi R/2$. On the $S^1$, $y$=0 is
identified with $y = \pi R$ ($Z^\prime_2$ symmetry) while
$y = \pm \pi R$/2 are identified with each other ($Z_2$
symmetry).

As is common in models of this type, we assume that the fermions are
located in the 4-D brane at $y = 0$ while the gauge bosons and the
scalars are allowed to travel in the bulk. The discrete $Z_2$ and
$Z^\prime_2$ symmetries, which we refer to as $P$ and $P^\prime$,
permit the expansion of any 5-D field $\phi$ in the following mode
expansions according to whether they are even or odd under ($P,P^\prime$):
\[
\begin{array}{cccc}
\phi_{++}(y) &=& \sqrt{2 \over \pi R} \sum_{n=0}^{\infty}
\phi^{(2n)}_{++}\cos{2 n y \over R}; & ~~~~M_n = \frac{2n}{R} \nonumber\\
\phi_{-+}(y) &=& \sqrt{2 \over \pi R} \sum_{n=0}^{\infty}
\phi^{(2n+1)}_{-+}\sin{(2 n +1) y \over R}; & ~~~~M_n =
\frac{2n+1}{R}\nonumber\\
\phi_{+-}(y) &=& \sqrt{2 \over \pi R} \sum_{n=0}^{\infty}
\phi^{(2n+1)}_{+-}\cos{(2 n +1) y \over R}; & ~~~~M_n = \frac{2n+1}{R}
\nonumber\\
\phi_{--}(y) &=& \sqrt{2 \over \pi R} \sum_{n=0}^{\infty}
\phi^{(2n+2)}_{--}\sin{(2 n +2) y \over R}; & ~~~~M_n = \frac{2n+2}{R}
\nonumber 
\end{array}
\]

Here $n = 0,1,2, \ldots$ and we have suppressed the $x^\mu$
dependence.  The behaviour of the fields under $P$ and $P^\prime$ can
be read off from the subscripts in the left hand side above. For
example, $\phi_{-+}$ is odd under $P$ and even under $P^\prime$. We
have also listed the masses of the different modes. Notice that only
the $\phi_{++}$ field can have a massless mode. One of the prime
motivations of these higher dimensional $SU(5)$ models is to ensure
doublet-triplet splitting within the {\bf 5} scalar multiplet of
$SU(5)$ and to ensure that from within the adjoint representation
({\bf 24}) of the gauge bosons only the $SU(3)_c \times SU(2)_L \times
U(1)_Y$ gauge bosons remain massless. Both can be achieved by
ascribing $P, P^\prime$ parities of $++$ to the (1,2,1/2) submultiplet
in the {\bf 5} while the remaining (3,1,--1/3) states carries $+-$
parity\footnote{Here we are using the decomposition of the $SU(5)$
multiplets under $ SU(3)_c \times SU(2)_L \times U(1)_Y$.}. For the
sake of completeness, the decomposition of the $SU(5)$ representations
upto {\bf 75} are listed in Table (\ref{parity}) \cite{szz}.  Since
$P, P^\prime$ commute with the Standard Model (SM) gauge symmetry
$SU(3)_c \times SU(2)_L \times U(1)_Y$, the $Z_2 \times Z^\prime_2$
parities for the members of the higher $SU(5)$ multiplets can be built
up from this assignment for the scalars in the fundamental
representation\footnote{In principle, one can assign arbitrary $Z_2
\times Z^\prime_2$ parities to the submultiplets of the higher SU(5)
representations. However, to reduce adhocness, here we work with the
{\em ansatz} that once we assign the parities for the fundamental
representation, those for the submultiplets of higher $SU(5)$
representations are determined by group theoretic relationships.}.
These parities have been indicated in Table (\ref{parity}). In this
way we can also assure that the (1,1,1)+(1,3,0)+(8,1,0) multiplets of
${ \bf 24}$ remain massless, breaking $SU(5)$ symmetry below the
compactification scale $1/R \equiv M_X$.

\begin{table}[ht]
\[
\begin{array}{|rcl|}
\hline 
SU(5) \supset && SU(3)_c \times SU(2)_L \times U(1)_Y  \nonumber \\
\hline
{\bf 5}    \supset && ( 1,2,1/2)_{\bf+~+} + (3,1,-1/3)_{\bf+~-} \nonumber\\  
{\bf \overline{5}}    \supset && ( 1,2,-1/2)_{\bf+~+} 
+ (\overline{3},1, 1/3)_{\bf+~-} \nonumber\\  
{\bf 10}    \supset && ( 1,1,1)_{\bf+~+} + (\overline{3},1,-2/3)_{\bf+~+} 
+(3,2,1/6)_{\bf+~-} \nonumber\\
{\bf 15}    \supset && ( 1,3,1)_{\bf+~+} + (3,2,1/6)_{\bf+~-} 
+(6,1,-2/3)_{\bf+~+} \nonumber\\
{\bf 24}    \supset && ( 1,1,0)_{\bf+~+} + (1,3,0)_{\bf+~+} 
+(3,2,-5/6)_{\bf+~-} +(\overline{3},2,5/6)_{\bf+~-}+(8,1,0)_{\bf+~+} \nonumber\\  
{\bf 35} \supset && ( 1,4,-3/2)_{\bf+~+} + (\overline{3},3,-2/3)_{\bf+~-} 
+(\overline{6},2,1/6)_{\bf+~+} 
+(\overline{10},1,1)_{\bf+~-}\nonumber\\  
{\bf 40}    \supset && ( 1,2,-3/2)_{\bf+~+} + (3,2,1/6)_{\bf+~+} 
+(\overline{3},1,-2/3)_{\bf+~-}
+(\overline{3},3,-2/3)_{\bf+~-} \nonumber\\
&& +(8,1,1)_{\bf+~-}+(\overline{6},2,1/6)_{\bf+~+}  
\nonumber\\
{\bf 45}    \supset && ( 1,2,1/2)_{\bf+~+} + (3,1,-1/3)_{\bf+~-} 
+(3,3,-1/3)_{\bf+~-}
+(\overline{3},1,4/3)_{\bf+~-} \nonumber\\
&& +(\overline{3},2,-7/6)_{\bf+~+}
+(\overline{6},1,-1/3)_{\bf+~-}+(8,2,1/2)_{\bf+~+}  
\nonumber\\
{\bf 50}    \supset && ( 1,1,-2)_{\bf+~+} + (3,1,-1/3)_{\bf+~+} 
+(\overline{3},2,-7/6)_{\bf+~-}
+(\overline{6},3,-1/3)_{\bf+~-} \nonumber\\
&& +(6,1,4/3)_{\bf+~+}
+(8,2,1/2)_{\bf+~-}  \nonumber\\
{\bf 70}    \supset && ( 1,2,1/2)_{\bf+~+} + (1,4,1/2)_{\bf+~+} 
+(3,1,-1/3)_{\bf+~-}
+(3,3,-1/3)_{\bf+~-} \nonumber\\
&& +(\overline{3},3,4/3)_{\bf+~-}
+(6,2,-7/6)_{\bf+~+}  +(8,2,1/2)_{\bf+~+}  +(15,1,-1/3)_{\bf+~-}  
\nonumber\\
{\bf 70^\prime}    \supset && ( 1,5,-2)_{\bf+~+} 
+ (\overline{3},4,-7/6)_{\bf+~-} 
+(\overline{6},3,-1/3)_{\bf+~+}
+(\overline{10},2,1/2)_{\bf+~-} \nonumber\\ 
&& +(\overline{15},1,4/3)_{\bf+~+} 
\nonumber\\
{\bf 75} \supset && ( 1,1,0)_{\bf+~+} 
+ (3,1,5/3)_{\bf+~+} 
+(3,2,-5/6)_{\bf+~-}
+(\overline{3},1,5/3)_{\bf+~+} \nonumber\\ 
&& 
+(\overline{3},2,5/6)_{\bf+~-} 
+(\overline{6},2,-5/6)_{\bf+~-} 
+(6,2,5/6)_{\bf+~-} 
+(8,1,0)_{\bf+~+} \nonumber\\ 
&& +(8,3,0)_{\bf+~+} 
\nonumber\\
\hline
\end{array}
\]
\caption{The $SU(3)_c \times SU(2)_L \times U(1)_Y$ contents of the different
$SU(5)$ representations. Also shown are the $P$ and $P^\prime$
assignments.}
\label{parity}
\end{table}

Above $M_X$, the mass scale of the non-SM gauge bosons, $X,Y$, and
that of the colour triplet scalars in the {\bf 5} representation,
$SU(5)$ symmetry is unbroken. This scale is determined in our analysis
by the unification of the three SM gauge couplings. We assume one
intermediate scale, $M_I$, such that all scalars which are permitted
to have a zero mode, excepting the SM scalar doublet, pick up a mass at
this scale. We include their contributions to the beta functions of
the one loop renormalization group equations (RGE) and solve for both
the intermediate scale $M_I$ and the unification scale $M_X$. The beta
function coefficients are given by:

\bea
b_i &=&
\pmatrix{0 \cr -22/3 \cr -11 } 
+ n_f \pmatrix{ 4/3 \cr 4/3 \cr 4/3 }
+ T_s^i/3 \label{beta}
\eea
We take $n_f = 3$. The above expression assumes that the scalar fields
are complex. For real scalar fields one has to use $T_s^i/6$ in
Eqn. (\ref{beta}). The $T_s^i$ for the light scalar submultiplets of
the different $SU(5)$ representations upto {\bf 75} are listed in Table
(\ref{contrib}).
\begin{table}
\begin{tabular}{|ccccc|}
\hline 
 R  & light~scalar~multiplets  & $T^3_s$ & $T^2_s$ & $T^1_s$ \\
\hline 
{\bf 5}  & (1,2,1/2) & 0 & 1/2 & 3/10 \\
{\bf 10} & (1,1,1)+($\overline{3}$, 1,- 2/3) & 1/2 & 0 & 7/5 \\
{\bf 15} & (1,3,1)+(6,1,-2/3)&5/2 &2 & 17/5 \\
{\bf 24} & (1,3,0)+(8,1,0)&3 &2 &0 \\
{\bf 35} & (1,4,-3/2)+($\overline{6}$,2,1/6)&5  & 8 & 28/5 \\
{\bf 40} & (1,2,-3/2)+(3,2,1/6)+($\overline{6}$,2,1/6) & 6 & 5& 3 \\
{\bf 45} & (1,2,1/2)+($\overline{3}$,2,-7/6)+(8,2,1/2) & 7 & 6 & 38/5 \\
{\bf 50} & (1,1,-2)+(3,1,-1/3)+ ($\overline{6}$,3,-1/3)+(6,1,4/3) 
& 21/2 & 3  & 51/5 \\
{\bf 70} & (1,2,1/2)+(1,4,1/2)+(6,2,-7/6)+(8,2,1/2) & 11 & 25/2 & 131/10 \\
${\bf 70^\prime }$& (1,5,-2)+($\overline{6}$,3,-1/3)+($\overline{15}$,1,4/3) 
& 25 & 22  & 146/5 \\
{\bf 75} & (3,1,5/3)+($\overline{3}$,1,-5/3)+(8,1,0)+(8,3,0) & 13 & 16 & 10\\
\hline
\end{tabular}  
\caption{The contributions to the $\beta$-functions from the
light members of the different $SU(5)$ representations upto {\bf 75}.}
\label{contrib}
\end{table}
Defining $m_{k,l}= \ln({m_k/m_l})$ and $b^i_{k,l}$ to be the
$\beta$ coefficients governing evolution in the range $m_k
\leftrightarrow m_l$, we get the following three solutions of the RGE.
\begin{equation} 
2 \pi \alpha^{-1}_i(M_Z) = 2 \pi \alpha^{-1}_{X}
+b^i_{X,I} m_{X,I} + b^i_{I,Z} m_{I,Z} 
\label{eqns} 
\end{equation}
Using the values of couplings at the low energy scale $M_Z$
\begin{equation}
\alpha_1(M_Z)=0.01688,~~~\alpha_2(M_Z)=0.03322,
~~~\alpha_3(M_Z)=0.117 \label{values}
\end{equation}
we solve the three equations in Eqn. (\ref{eqns}). First, we
present a simple illustrative example below.
\section{Simple example}
\begin{figure}[htb] \begin{center} \epsfxsize=11cm \epsfysize=11cm
\mbox{\hskip 0in}\epsfbox{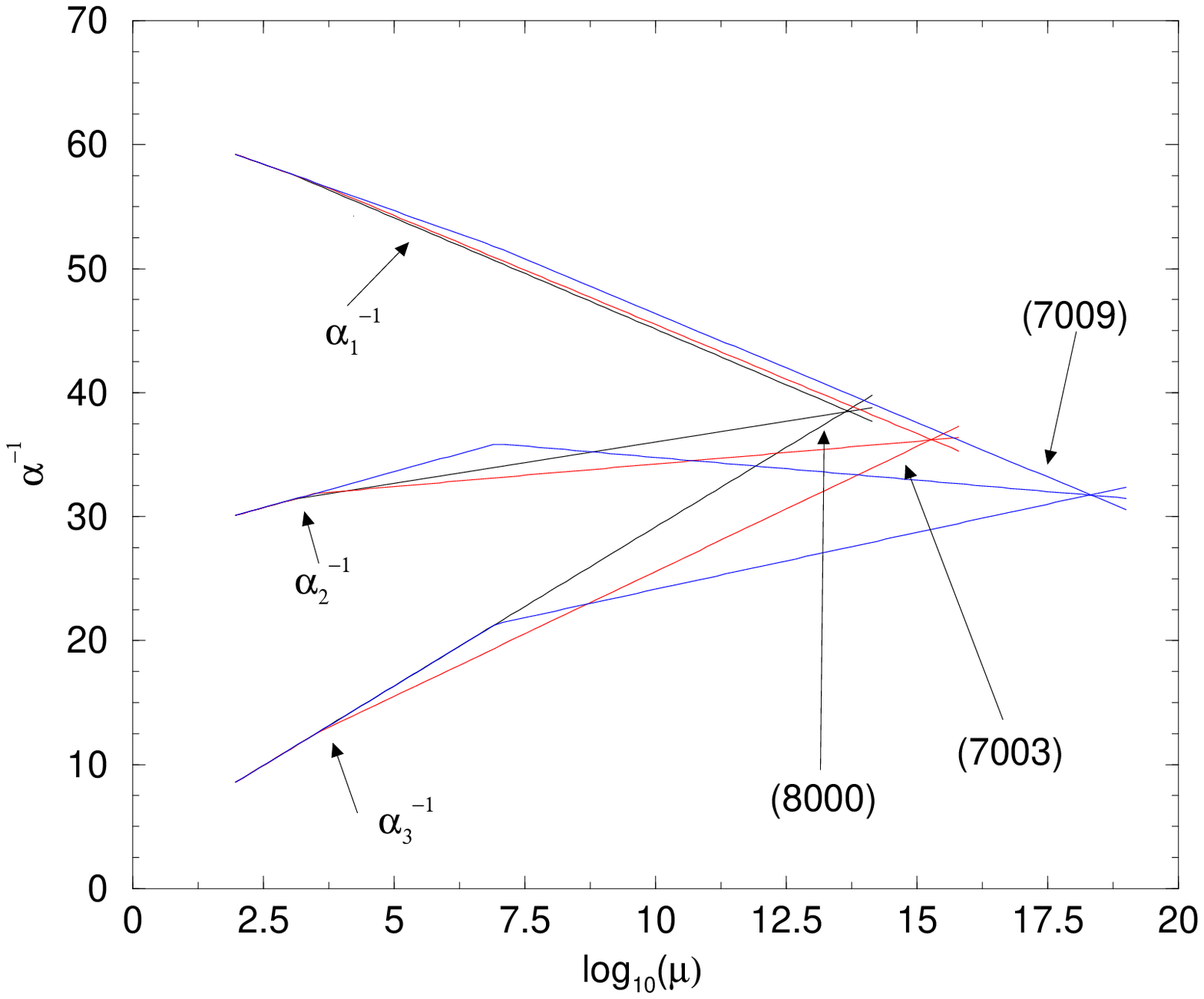}
\caption{Gauge unification in various models. Labels of cases
are $(n_5,n_{10},n_{15},n_{24})$. As a first approximation we
have used one intermediate scale which is given by the mass scale of 
extra scalars allowed by $S^1/Z_2 \times Z^\prime_2$ compactifications.} 
\label{fig1} 
\end{center}
\end{figure}

Because the GUT symmetry is broken via orbifolding, let us
consider the case where there are only {\bf 5}-plets of $SU(5)$ Higgs
scalars at the
unification scale and assume that there are $n_5$ of them. Then
compactification allows only doublets to be light and not their
triplet partners. In this case the $\beta$ coefficients are given by,
\bea 
b^i_{X,I}=\pmatrix{41/10 \cr -19/6 \cr
-7 } + {n_5 \over 3} \pmatrix{3/10 \cr 1/2 \cr 0} 
\label{case1} 
\eea
Solving Eqn. (\ref{eqns}) we obtain
\begin{equation}
\alpha^{-1}_{X}=38.53, m_{I,Z}=26.98-194.75/n_5,
m_{X,I}=194.75/n_5
\end{equation}
Because $m_{I,Z} \ge 0$ we obtain $n_5 \ge 8$.
For the case of $n_5=8$ we get,
\begin{equation}   
M_I = 1.39~ {\rm TeV}, M_{X} =5.0 \times 10^{10}~ {\rm TeV}.
\end{equation}
The GUT scale $M_X$ is rather low but it is consistent with proton
decay because of the existence of $Z_2 \times Z^\prime_2$ parity.
Note that the intermediate scale is in a very attractive region
phenomenologically. Eight Higgs doublets can be degenerate at this
scale of 1.3-1.4 TeV. They may play an important role in the fermion mass
puzzle. Further, the scale $M_X\simeq10^{10}$ TeV is interesting from the
point of view of the see-saw mechanism. The unification pattern is
shown in Fig. (\ref{fig1}).

\section{More general cases}
We now turn to the more general possibility where scalars in higher
representations of $SU(5)$ are present. The light scalars of all
$SU(5)$ multiplets upto {\bf 75} and their contributions to the beta
coefficients are listed in Table (\ref{contrib}).
\subsection{Small number of representations and low intermediate scales}
We consider upto the {\bf 75} dimensional representation of $SU(5)$
and demand that the threshold scale, $M_I$, be less than 10 TeV. For
the sake of economy, we also consider only those solutions where for
every representation $R$, the number $n_R$ is either 0 or 1.  If we do
not put any restriction on the number of representations, but maintain
that $n_R$ be zero or unity only, then we get 43 different solutions.
In Table (\ref{su5l}) we list those solutions for which not more than
two $n_R$ are non-zero.
\begin{table}[ht]
\begin{center}
\begin{tabular}{|ccc|}
\hline 
 Representations & $M_I$ & $M_X$ \\
 with $n_R=1$ & (TeV) & (TeV) \\
\hline
 {\bf 35} & 0.223 & 1.61 $\times 10^{11}$\\
 {\bf 5,35} & 5.70 & 1.38 $\times 10^{11}$\\
 {\bf 24,35} & 0.905 & 5.50 $\times 10^{11}$\\
 {\bf 35,40} & 3.61 & 1.84 $\times 10^{12}$\\
 {\bf 45,75} & 3.61 & 1.84 $\times 10^{11}$\\
\hline 
\end{tabular}
\end{center}
\caption{$SU(5)$ repesentations of scalars whose light members ensure
coupling constant unification. The unification scale, $M_X$,  and the
intermediate scale, $M_I$, are also given.}
\label{su5l}
\end{table}

Though the intermediate scales, $M_I$, and the unification scales,
$M_X$, in the last two cases are the same, the value of $\alpha_{X}$
turns out to be 0.037 and 0.710, respectively.

Let us explain one case in more detail. Let there be only 
{\bf 35}-plets of $SU(5)$ at the unification scale and assume that there 
are $n_{35}$ of them. Then compactification allows only (1,4,--3/2)
+ ($\overline{6}$,2,1/6) fields at low energy. In this case the $\beta_i$ 
coefficients are given by,
\begin{equation}
b^i_{X,I}=\pmatrix{41/10 \cr -19/6 \cr
-7 } + {n_{35} \over 3} \pmatrix{28/5 \cr 8 \cr 5} 
\label{case2} 
\end{equation}
Solving the RGE we obtain
\begin{equation}
\alpha^{-1}_{X}=32.72, m_{I,Z}=28.20-27.3073/n_{35},
m_{X,I}=27.3073/n_{35}
\end{equation}
Because $m_{I,Z} \ge 0$ we obtain $n_{35} \ge 1$.
For the case of $n_{35}=1$ we get,
\begin{equation}   
M_I = 0.223~ {\rm TeV},M_{X} =1.6 \times 10^{11}~ {\rm TeV}.
\end{equation}

\subsection{Low dimensional representations only}
Another alternative which we examine is by restricting to $SU(5)$
representations upto {\bf 24} subject further to the requirements $n_5
< 8, n_{10} < 5, n_{15} < 5 , n_{24} < 5$. Then we get the results
given in Table (\ref{lowdim}) when we impose $M_I < 70$ TeV. The
unification patterns of the gauge couplings for a few sample cases are
shown in Fig. (1).
\begin{table}[ht]
\begin{center}
\begin{tabular}{|cccccc|}
\hline 
$n_5$& $n_{10}$ & $n_{15}$ & $n_{24}$ & $M_I~{\rm(TeV)}$ & $M_X~{\rm(TeV)}$ \\
\hline
 8 & 0 & 0 & 0 & 1.678 & $4.66 \times 10^{10}$ \\
 7 & 0 & 0 & 1 & 0.223 & $1.61 \times 10^{11}$ \\
 8 & 0 & 0 & 1 & 5.696 & $1.38 \times 10^{11}$ \\
 7 & 0 & 0 & 2 & 0.905 & $5.49 \times 10^{11}$ \\
 8 & 0 & 0 & 2 & 19.08 & $4.05 \times 10^{11}$ \\
 7 & 0 & 0 & 3 & 3.607 & $1.84 \times 10^{12}$ \\
 8 & 0 & 0 & 3 & 63.08 & $1.17 \times 10^{12}$ \\
 6 & 0 & 0 & 4 & 0.407 & $1.33 \times 10^{13}$ \\
 7 & 0 & 0 & 4 & 14.12 & $6.08 \times 10^{12}$ \\
 6 & 0 & 0 & 5 & 0.199 & $5.22 \times 10^{13}$ \\
 7 & 0 & 0 & 5 & 54.36 & $1.97 \times 10^{13}$ \\
 8 & 1 & 0 & 5 & 0.407 & $1.33 \times 10^{13}$ \\
 8 & 0 & 1 & 5 & 0.407 & $1.33 \times 10^{13}$ \\
\hline 
\end{tabular}
\end{center}
\caption{Various $SU(5)$ scrnarios which gives low intermediate scales
upto 70 TeV}
\label{lowdim}
\end{table}

\section{Remarks about $SO(10)$}

It might be of interest to extend this analysis to grand unification
groups of higher rank. It is readily seen that the solutions will
become more difficult to come by. For example, we give in Table
(\ref{so10}) the $SU(5)$ contents of the $SO(10)$ representations upto
{\bf 126}. Notice, that the inclusion of a single {\bf 126} of
$SO(10)$ is equivalent to the simultaneous presence of {\bf
$\overline{\bf 5}$, 10, $\overline{\bf 15}$, 45, $\overline{\bf 50}$}
repesentations of $SU(5)$ and there is no flexibility of including the
$SU(5)$ representations individually.

\begin{center}
\begin{table}[ht]
\begin{tabular}{|ccccc|}
\hline 
 R  & SU(5)~components& $T^3_s$ & $T^2_s$ & $T^1_s$ \\
\hline 
{\bf 10} & ${\bf 5+ \overline{5}}$  & 0 & 1 & 3/5 \\
{\bf 16} & ${\bf 1+ \overline{5}+ 10}$ & 1/2 & 1/2 & 17/10 \\
{\bf 45} & ${\bf 1+10+ \overline{10} +24}$ & 4 & 2 & 14/5 \\
{\bf 120} & ${\bf 5+ \overline{5} + 10 +  \overline{10} 
+ 45 + \overline{45}}$ &15 &13 &93/5 \\
{\bf 126} & ${\bf 1+ \overline{5} + 10 + \overline{15} + 45 
+ \overline{50}}$ &41/2  & 23/2 & 229/10 \\
\hline
\end{tabular}  
\caption{The contributions to the $\beta$-functions from the
light members of the different SO(10) representations 
upto {\bf 126}.}
\label{so10}
\end{table}
\end{center}

If we permit all SO(10) representations upto {\bf 126} and consider no
more than upto 8 of any single representation then we find just two
allowed solutions:\\

1) $n_{10} = 8$, other $n_i = 0 \Rightarrow M_I$ = 1.68 TeV, $M_X$ =
   4.67 $\times 10^{10}$ TeV\\

2) $n_{10} = 6$, $n_{16} = 1$, other $n_i = 0 \Rightarrow M_I$ = 295
   TeV, $M_X$ = 7.99 $\times 10^{9}$ TeV\\

\section{Conclusions and Discussion}

In this work, we have examined the light scalar modes that survive
when a 5-dimensional $SU(5)$ model reduces to the 4-dimensional SM 
through the orbifold compactification route. The scalars which
are permitted to have zero modes are assumed to pick up a mass at some
scale $M_I$ intermediate between the electroweak and Planck
scales. They contribute to the beta coefficients in the $M_I < \mu <
M_X$ regime. The compactification scale, $M_X$, above which $SU(5)$ is
unbroken, is determined by the unification of the gauge
couplings. This analysis also determines $M_I$. We identify solutions
for which $M_I$ is in an interesting phenomenological range and can be
probed at the next generation colliders.  This analysis is somewhat
similar in spirit to the approach chosen for supersymmetric-GUTs where the
SUSY scale is fixed by gauge unification.

It is seen from Table (\ref{contrib}) that $P$ is (+)
for all the multiplets. Thus it does not play any role in the
present analysis. However, we would like to keep the option of
generalizing this method to the supersymmetric case where
$P$ has a non-trivial role. 

The scale of degenerate scalars, $M_I$, should be treated as an
approximate one in the sense that in reality some spread in the masses
around it can be expected. The standard model doublet has a mass at
the electroweak scale. This should not be viewed as an unnatural fine
tuning as some relevant Yukawa couplings can be of order
$10^{-2}-10^{-3}$.

When the unification symmetry is assumed to be of higher rank, then
the number of acceptable solutions reduces dramatically. For $SO(10)$
just two solutions can be obtained, so long as we stick to the low
dimensional representations of the symmetry group.

At first sight it might seem that we are introducing too many scalar
degrees of freedom. However, this appears  less dramatic 
when we compare it to the Minimal Supersymmetric Standard
Model  where too a large number of scalars are required.  \\

{\large{\bf Acknowledgements:}} This work was done while both authors
were visiting the Abdus Salam International Centre for Theoretical
Physics, Italy. They are grateful to the High Energy Section and the
Associateship Office of the Centre for hospitality. The research of AR
is supported by CSIR, India.

\newpage

\end{document}